\documentclass[twocolumn,preprintnumbers,nofootinbib,showpacs,amsmath,amssymb,prd,superscriptaddress]{revtex4-1}
\usepackage{amsmath}
\usepackage{color}
\usepackage{graphicx}
\usepackage{amssymb}
\usepackage{bbm}
\usepackage{amsfonts}
\usepackage[%pdfstartview=FitH,
       colorlinks=true,
      filecolor=black,
      linkcolor=blue,
      citecolor=red,
      urlcolor=blue,
       linktocpage=true,
        pdfstartview=FitH,
        pdfpagemode=None,
        bookmarksopen=true,
            ]{hyperref}

\def\I{I }
\def\sc{$\Sigma_c$ }

\def\0{{(0)}}
\def\1{{(1)}}
\def\2{{(2)}}
\def\3O{O(\partial^3)}
\def\gv{{\gamma_v}}

\def\l{{\lambda}}
\def\a{{\alpha}}

\def\g{{\gamma}}
\def\de{{\delta}}

\def\t{\tau}
\def\e{{\epsilon}}
\def\ep{\epsilon}

\def\D{\partial}
\def\Oo{\Omega}

\def\p{\partial}

\def\D{\partial}
\def\d{\mathrm{d}}

\newcommand{\bfl}[1]{\mbox{\boldmath{$#1$}}}
\newcommand{\bml}[1]{\mbox{\scriptsize\boldmath{$#1$}}}

\def\bmell{{\bml{\ell}}}
\def\hk{{\bf k}}
\def\hell{{\bfl \ell}}
\def\hn{{\bf n}}
\def\hu{{\bf u}}
\def\hm{{\bf m}}

\def\hT{T}
\def\hhT{\hat{T}}

\def\hhm{m}
\def\ii{\!\!\!}

\def\tg{\tilde g}
\def\tx{\tilde x}
\def\tr{\tilde r}
\def\tu{\tilde u}
\def\ts{\tilde s}
\def\tT{\tilde T}

\def\erho{\mathrm{e}}
\def\pp{\mathrm{p}}
\def\Pe{\mathrm{P}}
\def\Ha{\mathrm{H}}
\def\pp{\mathbbm{p}}
\def\tpp{\tilde{\pp}}
\def\Pe{\mathbb{P}}
\def\Ha{\mathbb{H}}

\def\Pij{\mathrm{P}}
\def\erho{\mathbbm{e}}

\def\K{\mathcal{K}}

\def\nn{{\nonumber}}
\renewcommand{\(}{\left(}
\renewcommand{\)}{\right)}

\begin{document}

\title{\bf \Large Petrov type \I Spacetime and Dual Relativistic Fluids}
\author{Rong-Gen Cai}
\email[]{cairg@itp.ac.cn}
\affiliation{State Key Laboratory of Theoretical Physics, Institute of Theoretical Physics,
Chinese Academy of Sciences, Beijing 100190, People's Republic of China}
\author{Qing Yang}
\email[]{yangqing@itp.ac.cn}
\affiliation{State Key Laboratory of Theoretical Physics, Institute of Theoretical Physics,
Chinese Academy of Sciences, Beijing 100190, People's Republic of China}
\author{Yun-Long Zhang}
\email[]{zhangyl@itp.ac.cn}
\affiliation{State Key Laboratory of Theoretical Physics, Institute of Theoretical Physics,
Chinese Academy of Sciences, Beijing 100190, People's Republic of China}
\affiliation{Mathematical Sciences and STAG research centre,
University of Southampton, Southampton SO17 1BJ, United Kingdom}

\vspace{-20pt}
\begin{abstract}
The Petrov type \I condition for the solutions of vacuum Einstein equations in both of the non-relativistic and relativistic hydrodynamic expansions is checked. We show that it holds up to the third order of the non-relativistic hydrodynamic expansion parameter, but it is violated at the fourth order even if we choose a general frame. On the other hand, it is found that the condition holds at least up to the second order of the derivative expansion parameter. Turn the logic around, through imposing the Petrov type \I condition and Hamiltonian constraint on a finite cutoff surface, we show that the stress tensor of the relativistic fluid can be recovered with correct first order and second order transport coefficients dual to the solutions of vacuum Einstein equations.
\end{abstract}
% 04.50.Gh}
\date{ August 28, 2014}
%\date{\today}
\maketitle

%\newpage

%\begin{narrow}{0.6in}{0.6in}
%\tableofcontents
%\end{narrow}
%\allowdisplaybreaks

%\section{Introduction}
\emph{Introduction.} ---
The holographic duality between gravity and one lower dimensional fluid has attracted  much attention over the past years. There exist two kinds of prescriptions for the dual fluid. One is
the membrane paradigm which describes a fluid living on the stretched horizon of a black hole~\cite{Price:1986yy,Kovtun:2003wp,Gourgoulhon:2005ng,Eling:2009pb,Eling:2009sj},
and the other is the AdS/fluid duality which describes a certain conformal fluid living on the
anti-de Sitter (AdS) boundary~\cite{Policastro:2002se,Bhattacharyya:2008jc,VanRaamsdonk:2008fp,Haack:2008cp,Bhattacharyya:2008mz,Bhattacharyya:2008kq}.
It is expected that there exists some connection between the two descriptions~\cite{Iqbal:2008by,Nickel:2010pr,Eling:2011ct}.
This  motivates the authors in~\cite{Bredberg:2010ky} to consider the gravitational fluctuations confined inside a finite cutoff surface outside a horizon, and in this case the dual fluid lives on this hypersurface.  The Dirichlet condition on the cutoff surface and the regularity on the horizon are imposed. This procedure has also been
generalized to the asymptotically flat~\cite{Bredberg:2011jq,Bredberg:2011xw} and de Sitter~\cite{Anninos:2011zn} spacetimes.

The authors of \cite{Bredberg:2011jq} have shown that for every solution of the incompressible Navier-Stokes equations in $p+1$ dimensions, there exists a unique corresponding solution of  vacuum Einstein equations in $p+2$ dimensions. On the cutoff surface, the extrinsic curvature is given by the stress tensor of the Navier-Stokes fluid.  A systematical method to reconstruct the solution of vacuum Einstein gravity to an arbitrary order has been presented in both of the non-relativistic and relativistic hydrodynamic expansions~\cite{Compere:2011dx,Compere:2012mt,Eling:2012ni,Meyer:2013sva}.
It is interesting to note that, instead of imposing the regularity condition on the horizon,  imposing the Petrov type \I condition on a hypersurface in near-horizon limit is alternatively introduced in~\cite{Lysov:2011xx}. The Petrov type \I condition just gives $ {p(p+1)}/{2}$ constraints on the extrinsic curvature (or say, the Brown-York stress tensor $T_{ab}$ of the dual fluid), which leads to $p+1$ independent variables.  These variables are exactly the degrees of freedom of a fluid in $p+1$ dimensions. They have shown that combining the Petrov type \I condition with Hamiltonian and momentum constraints can lead to the incompressible Navier-Stokes equation for the dual fluid on the cutoff surface in the near-horizon limit. Some further generalizations and discussions can be seen in~\cite{Huang:2011he,Huang:2011kj,Zhang:2012uy,Wu:2013kqa,Wu:2013mda,Ling:2013kua,Lysov:2013jsa,Wu:2014fwa}.

Notice that if one considers the mathematically equivalent solution of vacuum Einstein equations in the non-relativistic hydrodynamic expansion with parameter $\ep$, the Petrov type \I condition holds up to order of $\ep^2$.
An interesting question is whether the solution of vacuum Einstein equations satisfies the Petrov type \I condition to higher orders.  It is found in \cite{Cai:2013uye} that the condition holds up to order $\ep^3$ and is broken at order $\ep^4$. However, those %Petrov type \I condition
violated terms contain only the third order terms of the derivative expansion parameter $\p$ %of relativistic fluid
if an improved frame is taken.  This motivates us to check the Petrov type \I condition for the solution of vacuum Einstein equations in the relativistic hydrodynamic expansion. It turns out that the condition indeed holds up to the second order of the derivative expansion parameter $\p$, by using the vacuum solution available to this order in \cite{Compere:2012mt}.
%This will be shown in the next section.

%\section{Petrov type \I condition in non-relativistic hydrodynamic expansion}
\emph{Petrov type \I spacetime in the non-relativistic hydrodynamic expansion.} ---
Let us start with the $p+2$ dimensional Rindler metric
\begin{align}\label{Rindler}
\d s^2=g^{(r)}_{\mu\nu}\d x^\mu \d x^\nu=-r \d\t^2+2\d\t \d r+\d x_i \d x^i,
\end{align}
where $x^\mu=(r, \t, x^i)$, and $i=1,2,..., p$.
A spacetime is at least Petrov type \I if for some choice of frame, %there exists a frame field at every point such that
$C_{(\bmell)i(\bmell)j}\equiv\hell^\mu \hm_i^{\;\a}\hell^\nu \hm_j^{\;\beta}C_{\mu\alpha\nu\beta}=0$ at each point \cite{Coley:2004jv,Coley:2007tp}. % of the spacetime  Here $C_{\mu\alpha\nu\beta}$ is the Weyl tensor of the spacetime,
Here $\hell$, $\hk$, $\hm_i$ are the $p+2$ Newman-Penrose-like vector fields which obey $\hell_\mu\hell^\mu = \hk_\mu \hk^\mu =0$, $\hell_\mu \hk^\mu=1$, $g_{\mu\nu}{\hm_i}^\mu{\hm_j}^\mu=\delta_{ij}$ and all other products vanish.
One can show that the whole Rindler spacetime (\ref{Rindler}) is Petrov type \I with the frame chosen as~\cite{Lysov:2011xx}
\begin{align}\label{frameRindler}
 \hm_i=\p_i, \;\;\sqrt{2}\hell =\p_0-\hn,\;\; \sqrt{2}\hk=-\p_0-\hn,
\end{align}
where $\p_0=\p_\tau/\sqrt{r}$ and $\hn=\sqrt{r}\p_r+\p_0$.

On a timelike hypersurface \sc at $r=r_c$  with a flat induced metric $\g_{ab} \d x^a \d x^b=-r_c \d\t^2+\d x_i \d x^i$, one can define the $p+1$ velocity $u^a=\gv(1,\, v^i)$, where $\gv$  is fixed through $\gamma_{ab}u^au^b=-1$. %${a,b}$ indices are raised by $\gamma_{ab}$, $i,j$ indices are raised by $\delta_{ij}$.
Introducing the other parameter $P$ and regarding $v^i$ and $P$ as slowly varying functions of $x^a=(\t,x^i)$,
one can consider the perturbations of the metric (\ref{Rindler}) in non-relativistic hydrodynamic limit \cite{Bhattacharyya:2008kq,Bredberg:2011jq} that $v_i\sim \p_i\sim  \e,~P \sim \p_\tau \sim \e^2$. The solution of vacuum Einstein equations to an arbitrary order of $\ep$ can be constructed through keeping the induced metric flat and demanding the regularity on the horizon~\cite{Compere:2011dx}.

In order to check whether the solution to higher orders in~\cite{Compere:2011dx} is Petrov type \I or not, we  consider a frame by adding higher order corrections to the zeroth order frame (\ref{frameRindler}) as
\vspace{-2pt}
\begin{align}\label{NP}
\sqrt{2}\,\hell=&\p_0- \hn'+\hell_{(\ep)}+\hell_{(\ep^2)}+O(\ep^3),\nn\\
\sqrt{2}\,\hk=&-\p_0- \hn'+\hk_{(\ep)}+\hk_{(\ep^2)}+O(\ep^3),\nn\\
\hm_{1}=&\hm'_1+\hm_{1(\ep)}+\hm_{1(\ep^2)}+O(\ep^3),\nn\\
\hm_{i'}=&\p_{i'}+\hm_{i'(\ep)}+\hm_{i'(\ep^2)}+O(\ep^3),
\end{align}
where $i',j'=2,...,p$, and the two zeroth order normalized spatial vectors are $\hn'=(\sin\theta)\hn-(\cos\theta )\hm_1$, $\hm'_1=(\cos\theta)\hn+(\sin\theta )\hm_1$. As there exists the rotational symmetry among the $\hm_i$ vectors,  this choice does not lose any generality. Putting them and the Wely tensors of the spacetime with higher order corrections \cite{Compere:2011dx} into $C_{(\bmell)i(\bmell)j}$, we find that up to $\ep^2$,
%{\cred
\begin{align}
4C_{(\bmell)1(\bmell)1}=\,&r^{-1}(\sin\theta-1)^2\p_1 v_1,\nn\\
4C_{(\bmell)1(\bmell)i'}=\,&\big[r^{-1}(\sin\theta-1)^2-3 r_c^{-1}(\sin^2\theta-1)\big]\p_{[1} v_{i']},\nn\\
4C_{(\bmell)i'(\bmell)j'}=\,&r^{-1}(\sin\theta-1)^2 \p_{(i'}v_{j')}.
\end{align}
%}
%where subscriptions $(\e)$ and $(\e^2)$ denote the order of the non-relativistic expansion.
If demanding $C_{(\bmell)i(\bmell)j}$ vanishes at this order, $\sin\theta=1$ is the only consistent solution, which just gives the frame at the zeroth order (\ref{frameRindler}). % even include the higher order corrections of the frame.
Taking into account of this, the relevant possible choice of the first order corrections in (\ref{NP}) is
$\hell^\tau_{(\ep)}=0,~\hell^i_{(\ep)}=\l_{\ell}\sqrt{r} v^i,~\hm_{i(\ep)}^{~\tau}=\l_{m} v_i$,
where $\l_m$ and $\l_\ell$ are arbitrary functions of $r$ and $r_c$.
On the other hand, the orthogonal normalization condition of the vectors up to the first order of $\ep$ gives constraints that $\hm_{i(\ep)}^{~j}=0$ and $\hm_{i(\ep)}^{~\tau}-v_{i}/r_c=\delta_{ij}\hell_{(\ep)}^i$.
Putting them together we find that the non-vanishing terms in $C_{(\bmell)i(\bmell)j}$ first appear at order $\ep^4$,
\begin{align}\label{Petrovrij}
%\Pij^{(r)}_{ij}
4C_{(\bmell)i(\bmell)j}=&\l_{\ell} \,r_c^{-1}r\,\big[6\l_{\ell}v^k\omega_{k(i}v_{j)}+2v_{(i}\p^2 v_{j)}-4v^k\p_{(i}\omega_{j)k}\big]\nn\\&
+r_c^{-1}r\,\p^2\p_{(i}v_{j)}+O(\ep^5).
\end{align}
As all these terms in (\ref{Petrovrij}) are independent and only one free parameter $\l_{\ell}$ is left, it is impossible to make $C_{(\bmell)i(\bmell)j}$ in (\ref{Petrovrij}) vanish at $\ep^{4}$ for any choice of $\l_{\ell}$.
We may need to consider the possible higher order corrections to the velocity and pressure like $v_i\rightarrow v_i+\delta v_{i(\ep^3)}, P\rightarrow P+\delta P_{(\ep^4)}$,  but these corrections can be absorbed into the arbitrary functions $F^{_{(\ep^3)}}_i$ and $F^{_{(\ep^4)}}_\tau$ in the metric \cite{Compere:2011dx}, which do not make any contribution to $C_{(\bmell)i(\bmell)j}$ up to $\ep^4$.

Notice that by setting $\l_{\ell}=-r^{-1}$ and taking $r\rightarrow r_c$, one can recover the results in \cite{Cai:2013uye} that Petrov type \I condition is broken at $\e^4$, unless some additional physical conditions, such as the irrotational condition, are added. In particular, if setting $\l_{\ell}=0$ in (\ref{Petrovrij}), only the term $\p^2\p_{(i}v_{j)}$ with three derivatives is left. This seemingly implies that the Petrov type \I condition will be violated at the third order $\p^3$ of the derivative expansion. As no explicit solution of vacuum Einstein equations is available up to $\p^3$ in the literature,  therefore we are here not able to show whether the Petrov type \I condition holds at the third order and even arbitrary higher orders, although it is certainly of great interest to see this. In the following section, we will only consider the Petrov type \I condition of the solution of vacuum Einstein equations up to the second order in the derivative expansion.

%\section{Petrov type \I condition in relativistic hydrodynamic expansion}
\emph{Petrov type \I spacetime in the relativistic hydrodynamic expansion.} ---
Introduce the parameter $\pp=(r_c-r_h) ^{-1/2}$, which will turn out to be the pressure of the dual fluid, and $r_h$ is the location of the Rindler horizon of the equilibrium solution. %in the case of equilibrium state.
Then keeping the induced metric flat and demanding the regularity on the horizon,
regarding $u^a$ and $\pp$ as two slowly varying functions of $x^a$, one can obtain the solution of vacuum Einstein equations to an arbitrary order by using the derivative expansion. Up to the second order, the solution can be written as~\cite{Compere:2012mt}\vspace{-2pt}
\begin{align}\label{metric}
\d s^2=&g_{\mu\nu} \d x^\mu \d x^\nu= - 2\pp u_a \d x^a \d r+ g_{ab} \d x^a \d x^b,
\end{align}\vspace{-2pt}
where  $g_{ab}=g_{ab}^{\0}+g_{ab}^{\1}+g_{ab}^{\2}$, %+\3O$,
\vspace{-2pt}
\begin{align}
g_{ab}^{\0}=&-\pp^2(r-r_c)u_a u_b+\g_{ab},  \nn\\
g_{ab}^{\1}=&\,2\pp(r-r_c) \(u^c\p_c\!\ln\!\pp u_a u_b+2 a_{(a} u_{b)}\),\nn \\
g_{ab}^{(2)}=&\,2(r-r_c)\Big[(\K_{cd}\K^{cd})u_a u_b-2u_{(a} h_{b)}^c \p_d \K^d_{\;c}\nn\\&
- \K_a^{~c}\K_{cb}+ 2\K_{c(a}\Omega^c_{~b)}- 2 h_a^c h_b^d u^e\p_e \K_{cd}\Big]  \nn\\
&+ {\pp^2}(r-r_c)^2\Big\{ \big(\frac{1}{2}\K_{cd}\K^{cd}+ a_c a^c\big)u_a u_b\nn\\&
+2u_{(a}h_{b)}^c  \big[\p_d \K^d_{~c}-(\K_{cd}+\Oo_{cd})a^d\big]
-  \Omega_{ac}\Omega^{c}_{\; \, b}\Big\} \nn\\&
+ {\pp^4}(r-r_c)^3\big(\frac{1}{2} \Oo_{cd}\Oo^{cd}\big)u_a u_b.\label{gab2}
\end{align}
Here the transverse projector $h^{a}_{b}=\gamma^{a}_{b}+u^a u_b$, tensors $\K_{ab}= h_a^{c}h_b^{d}\p_{(c} u_{d)}$, $\Oo_{ab}=h_a^{c}h_b^{d}\p_{[c} u_{d]}$, acceleration $a^a=u^b\p_b u^a$.
%, and  $D=u^c \p_c$, $D_a^{\bot}=h_a^c \p_c$.
And the constraint equations are
\begin{align}\label{constraint}
\p_a u^a=&{2}{\pp}^{-1}\K_{ab}\K^{ab}+O(\p^3),\nn\\
a_a+h_a^b \p_b\!\ln\! \pp=&{2}{\pp}^{-1}h_a^c\p_b \K_c^b+O(\p^3).
\end{align}
Notice that $h^{a}_{b}$ can also be decomposed as $\hhm_i^{\;a}\hhm^i_{\;b}$, where
\begin{align}
\hhm_i^{\;a}=&\de_i^{\;a}+r_c^{-1/2}u_i\de_\t^a+(1+r_c^{1/2}\gv)^{-1}u_i u^j\de_j^a,
\end{align}
${a,b,...}$ and $i,j,...$ indices are raised (lowered) by $\gamma_{ab}$ and $\delta_{ij}$, respectively.
Denote $\hn$ being the spacelike unit normal of constant $r$ hypersurface, $\hu$ being the normalized $p+2$ velocity, and $\hm_i$ being the remaining orthonormal spatial vectors.
One then has
$g^{\mu\nu}=\hn^\mu \hn^\nu-\hu^\mu \hu^\nu+\de^{ij}\hm_i^{\;\mu}\hm_j^{\;\nu}$, where $\hn=\hn^r\p_r+\hn^a\p_a$, $\hu=\hu^a\p_a$, $\hm_i=\hm_i^{~a}\p_a$, and
\begin{align}\label{mi}
\hn^r=&\,\pp^{-1}\Big[1+\pp(r-r_c)\(\pp-2u^c\p_c\!\ln\!\pp\)\nn\\&~~~~~~~
+\big(-g^{\2}_{cd}+g^{\1}_{ac}g^{\1}_{bd}h^{ab}\big)u^c u^d\Big]^{1/2},\nn\\
\hn^a=&\,(\pp\hn^r)^{-1}\big[u^a+2\pp(r-r_c)a^a+
g^{\2}_{bc}u^{b}h^{ca}\big],\nn\\
\hu^a=&\,\hn^a,\qquad\hm_i^{~a}=\hhm_i^{~a}-\frac{1}{2} \hhm_i^{~b} g^{\2}_{bc}h^{ca}.
\end{align}
Further one can construct the two null vectors as
\begin{align} \label{frame}
 \sqrt{2}\hell^\mu =-\hn^\mu+ \hu^\mu,
 ~~\sqrt{2}\hk^\mu=-\hn^\mu - \hu^\mu,%=(-\hn^r,-2\hu^a),
\end{align}
 which obey $\hell_\mu\hk^\mu=1$ and all other products with $\hm_i^{~\mu}$ vanish. Along with the condition $g_{\mu\nu}\hm^\mu_i\hm^\nu_j=\delta_{ij}$ up to order $\p^2$, one can obtain the $p+2$ Newman-Penrose-like vector fields $\hell, \hk, \hm_i$ such that
\begin{align}
g_{\mu\nu}=\hell_\mu \hk_\nu +\hell_\nu \hk_\mu+\de_{ij}\hm^i_{\;\mu}\hm^j_{\;\nu}.
\end{align}
In this frame,
$\sqrt{2}\hell=\hn^r\p_r$ leads to the expression
\begin{align}\label{Clilj}
\Pij^{(r)}_{ij}\equiv2C_{(\bmell)i(\bmell)j}=\hm_i^{\;a} \hm_j^{\;b}\Pe^{(r)}_{ab},
\end{align}
where $\Pe^{{(r)}}_{ab}\ii\equiv \! \hn^r h_a^c \hn^r h_b^d C_{rcrd}$.\!
With the metric (\ref{metric}), we find
\begin{align}\label{Rrarb}
\!\Pe^{{(r)}}_{ab}=&-(\hn^{r})^{2}\big(\frac{1}{2}h_a^c h_b^d \partial_r^2 g^{(2)}_{cd}+ \pp^2\Oo_{ac}\Oo^c_{\;b}\big)+\3O,
\end{align}
and considering $g^{(2)}_{ab}$ in (\ref{gab2}), we conclude $\Pe^{{(r)}}_{ab}=O(\p^3)$, which also indicates $\Pij^{(r)}_{ij}=O(\p^3)$.  As a result, we have shown that the solution (\ref{metric}) of vacuum Einstein equations is Petrov type \I at each point up to the second order $\p^2$ in the derivative expansion.

%\subsection{On the cutoff hypersurface}
\emph{Petrov type I condition on the cutoff surface.} We can project the Weyl tensor on the hypersurface \sc and define $\Pij_{ij}\equiv 2C_{(\bmell)i(\bmell)j}|_{\Sigma_c}$. In \cite{Lysov:2011xx}, $\Pij_{ij}=0$ is named as Petrov type I condition %The latter
and $\Pij_{ij}$ can be rewritten in terms of the extrinsic curvature $K_{ab}$ of \sc by  employing the Gauss-Codazzi equations. Notice that $K_{ab}$ can be expressed in terms of the Brown-York stress tensor through $\hT_{ab}=2(K\gamma_{ab}-K_{ab})$.  We have
$\Pij_{ij}=m_i^{\;a} m_j^{\;b}\Pe_{ab}$
where
\begin{align}\label{petrovP}
4\Pe_{ab}=&\,h_a^m h_b^n\big[\({\hT}_{mc}{\hT}_{nd}- {\hT}_{mn}{\hT}_{cd}\)u^c u^d
- {\hT}_{mc}\hT^c_{~n}  \nn\\&%+4h_a^c h_b^d\big[
\qquad~~- 4u^c \partial_{c}{\hT}_{mn}+ 4 u^c\partial_{(m}\hT_{n)c}\big]\nn\\
&+{p^{-2}}\big[\hT({\hT}+p \,{\hT}_{cd}u^c u^d)+ 4 p\, u^c \partial_{c}{\hT}\big]h_{ab}.
\end{align}

With the bulk metric in (\ref{metric}), the dual stress tensor can be expanded in the following form
\begin{align}\label{Tab}
\hT_{ab} = \hT^{\0}_{ab}+\hT^{\1}_{ab}+\hT^{\2}_{ab}+O(\p^3),
\end{align}
and these terms are obtained in~\cite{Compere:2012mt} as
\vspace{-3pt}
\begin{align}
\hT^{\0}_{ab} &= \, \pp h_{ab}, \nn\\
\hT^{\1}_{ab} &= \, \zeta'(u^c\p_c\!\ln\! \pp) u_a u_b-2\eta\K_{ab},\nn\\
\hT^{\2}_{ab} &= \,  {\pp}^{-1}
\Big\{\big[ d_1 \K_{ab}\K^{ab} + d_2 \Omega_{ab}\Omega^{ab} + d_3(u^c\p_c\!\ln\! \pp )^2 \nn\\&
+d_4 u^c\p_c(u^d\p_d\!\ln\! \pp) +d_5 h^{cd}(\p_c \!\ln\! \pp)(\p_d\!\ln\! \pp)\big]u_a u_b \nn\\&
+ \big[ c_1 \K_{ac}\K^c_{~b}+\!c_2 \K_{c(a}\Omega^c_{~b)} +\! c_3 \Omega_{ac}\Omega^c_{~b} +\!c_4 h_a^c h_b^d\D_c\D_d \!\ln\! \pp\nn\\&
+ c_5 \K_{ab}(u^c\p_c\!\ln\! \pp)+ c_6 (h_a^c\p_c \!\ln\! \pp)(h_b^d\p_d\!\ln\! \pp)\big]\Big\}.\label{Tab2}
\end{align}
Here the first and second order transport coefficients are
\begin{align}
\zeta' =&\, 0 ,\quad \quad~ \eta = 1,\nn\\
d_1 =& -2,\quad d_2=d_3=d_4=d_5 = 0, \nn\\
c_1=&-2,\quad c_2 = c_3 = c_4 = c_5 = -c_6 = -4\, .\label{coefficients}
\end{align}
The momentum constraint $2G_{\mu b} \hn^{\mu}|_{\Sigma_c}=0$, which leads to the conservation of the stress tensor $\p^a \hT_{ab}=0$,  gives the constraint equations (\ref{constraint}), while
the Hamiltonian constraint $2G_{\mu\nu} \hn^{\mu}\hn^{\nu}|_{\Sigma_c}=0$  leads to $4\Ha\equiv p \hT_{ab}\hT^{ab}-\hT^2=0$, which can be viewed as the equation of state for the dual fluid. In addition,
one can show that the trace of the stress tensor satisfies $\hT=p\pp+O(\p^3)$. Putting the stress tensor (\ref{Tab}) into the expression (\ref{petrovP}), we then obtain $\Pe_{ab}= O(\p^3)$, which of course implies $\Pij_{ij}= O(\p^3)$. Thus we have shown again that the Petrov type \I condition $\Pij_{ij}=0$ is satisfied up to $\p^2$ by using the stress tensor of the dual relativistic fluid.

%\subsection{From Petrov type \I condition to dual relativistic fluid}
\emph{From Petrov type \I condition to dual relativistic fluid.}\label{Sec32}
In this subsection we turn the logic around. Assuming the Hamiltonian constraint and Petrov type \I condition on a finite cutoff surface, we will show that the stress tensor of the dual fluid can be fixed up to the second order of the derivative expansion, without using the details of the bulk metric. The resulting stress tensor exactly matches the one from the solution of vacuum Einstein equations.

Firstly, one can introduce an undetermined symmetric stress tensor $\hhT_{ab}$, and it satisfies %such that
$h_{a}^{\,b}\hhT_{bc} u^c=0$, where $u^a$ is regarded as the relativistic fluid velocity. Then the stress tensor can be decomposed as
$\hhT_{ab}=\,\erho u_au_b+\Pi_{ab}$,
where
\begin{align}
\erho\equiv \, \hhT_{ab} u^au^b,\quad
\Pi_{ab}\equiv \, h_a^ch_b^d\hhT_{cd}.
\end{align}
%{\cb We will also chosen the isotropy gauge where there is no higher order corrections to the pressure $\pp$.}
The Hamiltonian constraint becomes $\Ha=0$, where
\begin{align}
4\Ha\equiv&\, p\,(\erho^2+\Pi_{ab}\Pi^{ab})-\hhT^2,%\nn\\
\end{align}
and $\hhT=-\erho+\Pi_{ab} h^{ab}$. The Petrov type \I condition can be generalized as $\Pe_{ab}=0$, where
\begin{align}
\!\!\!4\Pe_{ab}\!\equiv&-\!\erho{\Pi}_{ab}-\!{\Pi}_{ac}{\Pi^c}_{b}-\!4h_a^ch_b^d (u^e\p_e{\Pi}_{cd})-\!{4 }\Pi_{(a}^{~\;c}h^d_{b)}\p_d u_c\nn\\
&-4\erho\K_{ab} +{p^{-2}}\big[\hhT({\hhT}+p \,\erho)+ 4 p\, u^c\p_c{\hhT}\big]h_{ab}.\!
\end{align}
Expanding the stress tensor in terms of the derivative expansion parameter $\p$ as
\begin{align}
   \erho   = &~ \erho^{(0)}+\erho^{(1)}+\erho^{(2)}+O(\p^3),\nn\\
\Pi_{ab}   = &~  \Pi_{ab}  ^{(0)}+ \Pi_{ab}  ^{(1)}+\Pi_{ab}   ^{(2)}+O(\p^3),
\end{align}
and we identify $\erho^{(0)}=0$, $\Pi_{ab}^{(0)}=\pp h_{ab}$ %having given the fact that
from the zeroth order Brown-York stress tensor in (\ref{Tab2}). % of the background Rindler spacetime, which gives form  % as the  of the background Rindler spacetime (\ref{Rindler}),
Then through
\vspace{-5pt}
\begin{align}
\Ha^{(1)}=&\,0\Rightarrow \erho^{(1)}=0,\\
\Pe_{ab}^{(1)}=&\,0\Rightarrow \Pi_{ab}^{(1)}=-2\K_{ab},
\end{align}
we can fix the stress tensor at the first order.  With these,
\begin{align}
&\Ha^{(2)}\!=0\Rightarrow \erho^{(2)}=-2 \pp^{-1} \K_{ab} \K^{ab},\\
&\Pe_{ab}^{(2)}\!=0\Rightarrow
\Pi_{ab}^{(2)}
=\pp^{-1}%\big[2\K_{ac}\K^c_{~b}-4\K_{c(a}\Oo^c_{~b)}+4h_a^c h_b^d D\K_{cd}\big].\nn\\
\big[-\!2\K_{ac}\K^c_{~b}- 4\K_{c(a}\Omega^c_{~b)}- 4\Omega_{ac}\Omega^c_{~b}\nn\\&\qquad\qquad\qquad\qquad\qquad
- 4h_a^c h_b^d\p_c\p_d \!\ln\!\pp- 4\K_{ab}(u^c\p_c\!\ln\! \pp)\nn\\&\qquad\qquad\qquad\qquad\qquad+ 4(h_a^c\p_c\! \ln\!\pp)(h_b^d\p_d\! \ln\! \pp)\big],
\end{align}
we can then fix the second order terms in the stress tensor.
In the above procedure we have chosen the isotropy gauge that there is no higher order corrections to the pressure $\pp$.
Thus, up to the second order, we obtain the total stress tensor of the dual relativistic fluid as
\begin{align}\label{Tcovab2}
\hhT_{ab}=&\erho^{\2} u_a u_b+ \pp h_{ab}+\Pi_{ab}^{(1)}+\Pi_{ab}^{(2)}.
\end{align}
%It can be shown that the above stress tensor which is obtained emply the Petrov type I condition,
It is identical to the Brown-York stress tensor in (\ref{Tab}) which is calculated from the whole metric (\ref{metric}).
%(\ref{Tab2}) by using some identities from the constraint equations (\ref{constraint})~\cite{Compere:2012mt}.
%the relation (\ref{rel4}).

%\section{Further discussions}\label{Sec3}

{\emph{Near-horizon expansion}}.
The relativistic hydrodynamic expansion can also be expressed in terms of the so-called alternative near-horizon expansion \cite{Compere:2012mt}.
 First take a Weyl rescaling $\d s^2\rightarrow\l^2\d s^2$, where the scaling parameter $\lambda$ is related to the cutoff $r_c$ as $\lambda=r_c^{1/2}$, then consider the relativistic hydrodynamic limit $\tx^a=\l x^a$ and the rescaled  metric $\d \ts^2 =\lambda^{2} \d s^2$, we can reach the metric in  the near-horizon expansion with parameter $\lambda$ as
\vspace{-2pt}
\begin{align}
\d \ts^2=&\tg_{\mu\nu}\d \tx^\mu \d \tx^\nu=-2\l^1 \tpp\tu_a \d \tx^a \d \tr \nn\\&
+\big(\tg^\0_{ab}+\l^1\tg^\1_{ab}+\l^2\tg^\2_{ab}\big)
\d\tx^a\d\tx^b,
\end{align}
where $\tg^\0_{ab},\tg^\1_{ab},\tg^\2_{ab}$ are just obtained from (\ref{gab2}) by mapping $(r_c, r, \pp, u^a)\rightarrow (\tr_c,\tr,\tpp,\tu^a)$, and setting  $\tr_c=1$.
With similar operation on the dual stress tensor in (\ref{Tab2}), the stress tensor $\tT_{ab}\d \tx^a\d \tx^b= \l^2 T_{ab}\d x^a\d x^b$ can be expressed as
$\tT_{ab}=\tT^\0_{ab}+\l^1\tT^\1_{ab}+\l^2\tT^\2_{ab}$.
Then all the previous discussions can be redone in the near-horizon expansion formulism. In particular,  the dynamic equations $\p^{\tilde a}\tT^\0_{ab}=0$ for a perfect relativistic fluid appear as an attractor, when $\lambda \rightarrow 0$.

\emph{Higher curvature gravity.} For asymptotically flat spacetime in higher curvature gravity, the effect of the Gauss-Bonnet term with coefficient $\a$ is studied in~\cite{Chirco:2011ex,Eling:2012xa}.
 With the solutions found there,  we find that $\Pij^{(r)}_{ij}=O(\p^3)$,
because the correction to the metric from the Gauss-Bonnet term appears only at order $\p^2$, and  the factor in front of the relevant terms $h_a^c h_b^d\delta g^{\2}_{cd}\propto\alpha(r-r_c)$, the latter
 will not make any contribution to (\ref{Rrarb}) up to order $\p^2$. Furthermore the dual stress tensor whose second order transport coefficients with the Gauss-Bonnet term correction can also be recovered through the Petrov type \I condition in the same way as in the present paper~\cite{Cai:2014sua}.

\emph{With a negative cosmological constant.}
In this case, the solution of Einstein equations will be asymptotically %anti-de Sitter (AdS)
AdS~\cite{Cai:2011xv,Kuperstein:2011fn,Brattan:2011my,Matsuo:2012pi,Kuperstein:2013hqa}, and we find that
$\Pij^{(r)}_{ij}\!=g^{rr}\hm_i^a\hm_j^bC_{rarb}\!\sim\!O(\p)$
under a similar frame as that in this paper.
%the relevant tensor $\Pe^{(r)}_{ab}=2\hn^r h_a^c \hn^r h_b^d C_{rcrd}\sim O(\p)$, which implies $\Pij^{(r)}_{ij}\sim O(\p)$ through (\ref{Clilj}). %This indicates that the frame used in this paper might not be suitable for the asymptotically AdS spacetime.
However, notice that the near-horizon limit $g^{rr}\rightarrow0$ leads to $\Pij^{(r)}_{ij}\rightarrow 0$, which indicates a close relation between the Petrov type \I condition and the membrane paradigm.
In particular, the ratio of shear viscosity over entropy density, $\eta/s|_{r_h}=[1-2(p+1)(p-2)\a]/4\pi$ at the horizon, can also be extracted through imposing Petrov type \I condition directly~\cite{CYZ}. Here $\alpha$ is the Gauss-Bonnet coefficient. And higher order transport coefficients can
also be obtained.
In addition, a so-called AdS/Ricci flat correspondence has been proposed recently in \cite{Caldarelli:2012hy,Caldarelli:2013aaa}, which can map asymptotically AdS black brane solutions~\cite{Bhattacharyya:2008mz} to asymptotically flat solutions~\cite{Compere:2012mt},
%. Using the correspondence,
and the dual stress tenor of Rindler fluid (\ref{Tab}) can be obtained exactly from the one of AdS fluid up to second order in derivative expansion. Thus, it would be interesting to see whether there exists a corresponding %Petrov type constraint
condition in the whole AdS or more general spacetime~\cite{Cai:2012vr,Pinzani-Fokeeva:2014cka}.

To summarize, we have shown that the whole spacetime is Petrov type \I for the solution of vacuum Einstein equations in the
non-relativistic hydrodynamic expansion up to the third order $\ep^3$, but it is violated at $\ep^4$ unless some additional
condition is imposed~\cite{Cai:2013uye}.
While in the relativistic hydrodynamic expansion, it holds at least up to the second order  $\p^2$.
As no explicit solution of vacuum Einstein equations is available up to $\p^3$ in the literature,
we are here not able to show whether the whole spacetime is Petrov type \I  at the third order and even arbitrary higher orders in the derivative expansion.
However, we can go a further step. The solution of vacuum Einstein equations up to $\ep^4$ in the non-relativistic hydrodynamic expansion can be captured by that in the relativistic hydrodynamic expansion up to $\p^3$~\cite{Compere:2011dx}.
If the whole spacetime is  Petrov type \I at order $\p^3$, it will also be Petrov type \I at $\ep^4$ in the non-relativistic hydrodynamic expansion. Our calculation in the non-relativistic expansion indicates that in general, the Petrov type \I condition will be violated at the third order $\p^3$ of the
relativistic hydrodynamical expansion parameter.

Turn the logic around, we have shown that imposing the Petrov type \I condition and Hamiltonian constraint on a finite cutoff surface, the stress tensor of the dual relativistic fluid can be fixed up to the second order of the derivative expansion. The resulting stress tensor identically matches the one calculated from the solution of vacuum Einstein equations.
As pointed out in \cite{Lysov:2011xx}, the Petrov type I condition is expected to be equivalent to the regularity condition on the future horizon of the spacetime, and it gives the constraint on the dual theory from gravity. We have indeed shown that imposing the Petrov type I condition is mathematically much simpler than imposing the regularity requirement, because one no longer needs to solve the perturbation equations of bulk gravity.
Notice that the boundary condition on the horizon has to imposed for the perturbations in the gravity/fluid duality, we therefore conclude that the Petrov type \I condition would indeed play an important role in this aspect.

%\vspace{-5pt}
%\section*{Acknowledgments}%\vspace{-5pt}
\emph{Acknowledgments.}---
The authors thank K. Skenderis and M. Taylor for their quite helpful comments and suggestions on this manuscript. This work is supported in part by National Natural Science Foundation of China (No.10821504, No.11035008, and No.11375247).
R. G. C. and Y. L. Z. would like to thank the Isaac Newton Institute for Mathematical Sciences, Cambridge, for support and hospitality during the programme ``Mathematics and Physics of the Holographic Principle''(September 16 - October 11, 2013), where some work on this paper was finished.
Y. L. Z. thanks helpful discussions and comments from M. M. Caldarelli, C. Eling, H. Liu, M. Rangamanni, as well as the support from China Scholarship Council (No.201204910341) and hospitality of Southampton University.


\begin{thebibliography}{99}
%\small
%\footnotesize
%\normalsize
%\cite{Price:1986yy}
\bibitem{Price:1986yy}
%\bibitem{Damour:1978cg}
  T.~Damour,
  %``Black Hole Eddy Currents,''
  \href{http://dx.doi.org/10.1103/PhysRevD.18.3598}{Phys.\ Rev.\ D {\bf 18}, 3598 (1978).}
  %%CITATION = PHRVA,D18,3598;%%
  %81 citations counted in INSPIRE as of 05 Mar 2014
  R.~H.~Price and K.~S.~Thorne,
  %%``Membrane Viewpoint On Black Holes: Properties And Evolution Of The Stretched Horizon,''
  \href{http://dx.doi.org/10.1103/PhysRevD.33.915}{Phys.\ Rev.\ D {\bf 33}, 915 (1986)}.
  %%CITATION = PHRVA,D33,915;%%
  %90 citations counted in INSPIRE as of 07 Sep 2013

%\cite{Kovtun:2003wp}
\bibitem{Kovtun:2003wp}
  P.~Kovtun, D.~T.~Son and A.~O.~Starinets,
  %%``Holography and hydrodynamics: Diffusion on stretched horizons,''
  \href{http://dx.doi.org/10.1088/1126-6708/2003/10/064}{JHEP {\bf 0310}, 064 (2003)}
  [\href{http://arxiv.org/abs/hep-th/0309213}{hep-th/0309213}].
  %%CITATION = HEP-TH/0309213;%%
  %353 citations counted in INSPIRE as of 07 Sep 2013


%\cite{Gourgoulhon:2005ng}
\bibitem{Gourgoulhon:2005ng}
  E.~Gourgoulhon and J.~L.~Jaramillo,
  %%``A 3+1 perspective on null hypersurfaces and isolated horizons,''
   \href{http://dx.doi.org/10.1016/j.physrep.2005.10.005}{Phys.\ Rept.\  {\bf 423}, 159 (2006)}
  [\href{http://arxiv.org/abs/gr-qc/0503113}{gr-qc/0503113}].
  %%CITATION = GR-QC/0503113;%%
  %77 citations counted in INSPIRE as of 07 Sep 2013


%\cite{Eling:2009pb}
\bibitem{Eling:2009pb}
  C.~Eling, I.~Fouxon and Y.~Oz,
  %``The Incompressible Navier-Stokes Equations From Membrane Dynamics,''
  \href{http://dx.doi.org/10.1016/j.physletb.2009.09.028}{Phys.\ Lett.\ B {\bf 680}, 496 (2009)}
  [\href{http://arxiv.org/abs/arXiv:0905.3638}{arXiv:0905.3638 [hep-th]}].
  %%CITATION = ARXIV:0905.3638;%%
  %41 citations counted in INSPIRE as of 07 Sep 2013

%\cite{Eling:2009sj}
\bibitem{Eling:2009sj}
  C.~Eling and Y.~Oz,
  %``Relativistic CFT Hydrodynamics from the Membrane Paradigm,''
  \href{http://dx.doi.org/10.1007/JHEP02(2010)069}{JHEP {\bf 1002}, 069 (2010)}
  [\href{http://arxiv.org/abs/arXiv:0906.4999}{arXiv:0906.4999 [hep-th]}].
  %%CITATION = ARXIV:0906.4999;%%
  %36 citations counted in INSPIRE as of 07 Sep 2013

%
%AdS
%
%%
%  AdS
%%%%%%%%%%%%%%%%%%%


%\cite{Policastro:2002se}
\bibitem{Policastro:2002se}
  G.~Policastro, D.~T.~Son and A.~O.~Starinets,
  %``From AdS / CFT correspondence to hydrodynamics,''
  \href{http://dx.doi.org/10.1088/1126-6708/2002/09/043}{JHEP {\bf 0209}, 043 (2002)}
  [\href{http://arxiv.org/abs/hep-th/0205052}{hep-th/0205052}].
  %%CITATION = HEP-TH/0205052;%%
  %383 citations counted in INSPIRE as of 07 Sep 2013


%\cite{Bhattacharyya:2008jc}
\bibitem{Bhattacharyya:2008jc}
  S.~Bhattacharyya, V.~E. Hubeny, S.~Minwalla and M.~Rangamani,
  %``Nonlinear Fluid Dynamics from Gravity,''
  \href{http://dx.doi.org/10.1088/1126-6708/2008/02/045}{JHEP {\bf 0802}, 045 (2008)}
  [\href{http://arxiv.org/abs/arXiv:0712.2456}{arXiv:0712.2456 [hep-th]}].
  %%CITATION = ARXIV:0712.2456;%%
  %400 citations counted in INSPIRE as of 05 Sep 2013

%\cite{VanRaamsdonk:2008fp}
\bibitem{VanRaamsdonk:2008fp}
  M.~Van Raamsdonk,
  %``Black Hole Dynamics From Atmospheric Science,''
\href{http://dx.doi.org/10.1088/1126-6708/2008/05/106}{JHEP {\bf 0805}, 106 (2008)}
[\href{http://arxiv.org/abs/arXiv:0802.3224}{arXiv:0802.3224 [hep-th]}].
  %%CITATION = ARXIV:0802.3224;%%
  %52 citations counted in INSPIRE as of 29 Aug 2013

%\cite{Haack:2008cp}
\bibitem{Haack:2008cp}
  M.~Haack and A.~Yarom,
  %``Nonlinear viscous hydrodynamics in various dimensions using AdS/CFT,''
  \href{http://dx.doi.org/10.1088/1126-6708/2008/10/063}{JHEP {\bf 0810}, 063 (2008)}
  [\href{http://arxiv.org/abs/arXiv:0806.4602}{arXiv:0806.4602 [hep-th]}].
  %%CITATION = ARXIV:0806.4602;%%
  %55 citations counted in INSPIRE as of 14 Feb 2014

%\cite{Bhattacharyya:2008mz}
\bibitem{Bhattacharyya:2008mz}
  S.~Bhattacharyya, R.~Loganayagam, I.~Mandal, S.~Minwalla and A.~Sharma,
  %``Conformal Nonlinear Fluid Dynamics from Gravity in Arbitrary Dimensions,''
  \href{http://dx.doi.org/10.1088/1126-6708/2008/12/116}{JHEP {\bf 0812}, 116 (2008)}
  [\href{http://arxiv.org/abs/arXiv:0809.4272}{arXiv:0809.4272 [hep-th]}].
  %%CITATION = ARXIV:0809.4272;%%
  %73 citations counted in INSPIRE as of 05 Sep 2013

%\cite{Bhattacharyya:2008kq}
\bibitem{Bhattacharyya:2008kq}
  S.~Bhattacharyya, S.~Minwalla and S.~R.~Wadia,
  %``The Incompressible Non-Relativistic Navier-Stokes Equation from Gravity,''
  \href{http://dx.doi.org/10.1088/1126-6708/2009/08/059}{JHEP {\bf 0908}, 059 (2009)}
  [\href{http://arxiv.org/abs/arXiv:0810.1545}{arXiv:0810.1545 [hep-th]}].
  %%CITATION = ARXIV:0810.1545;%%
  %82 citations counted in INSPIRE as of 05 Sep 2013


%%%%%%%%%%%%%%%%%%%%%%%%%%%%%%%%%%%%%%%%
%  Cutoff
%  Cutoff
%%%%%%%%%%%%%%%%%%%%%%%%%%%%%%%%%%%%%%

%\cite{Iqbal:2008by}
\bibitem{Iqbal:2008by}
  N.~Iqbal and H.~Liu,
  %``Universality of the hydrodynamic limit in AdS/CFT and the membrane paradigm,''
  \href{http://dx.doi.org/10.1103/PhysRevD.79.025023}{Phys.\ Rev.\ D {\bf 79}, 025023 (2009)}
  [\href{http://arxiv.org/abs/arXiv:0809.3808}{arXiv:0809.3808 [hep-th]}].
  %%CITATION = ARXIV:0809.3808;%%
  %207 citations counted in INSPIRE as of 07 Sep 2013

%\cite{Nickel:2010pr}
\bibitem{Nickel:2010pr}
  D.~Nickel and D.~T.~Son,
  %``Deconstructing holographic liquids,''
  \href{http://dx.doi.org/10.1088/1367-2630/13/7/075010}{New J.\ Phys.\  {\bf 13}, 075010 (2011)}
  [\href{http://arxiv.org/abs/arXiv:1009.3094}{arXiv:1009.3094 [hep-th]}].
  %%CITATION = ARXIV:1009.3094;%%
  %44 citations counted in INSPIRE as of 07 Sep 2013

%\cite{Eling:2011ct}
\bibitem{Eling:2011ct}
  C.~Eling and Y.~Oz,
  %``Holographic Screens and Transport Coefficients in the Fluid/Gravity Correspondence,''
  \href{http://dx.doi.org/10.1103/PhysRevLett.107.201602}{Phys.\ Rev.\ Lett.\  {\bf 107}, 201602 (2011)}
  [\href{http://arxiv.org/abs/arXiv:1107.2134}{arXiv:1107.2134 [hep-th]}].
  %%CITATION = ARXIV:1107.2134;%%
  %11 citations counted in INSPIRE as of 07 Sep 2013


%\cite{Bredberg:2010ky}
\bibitem{Bredberg:2010ky}
  I.~Bredberg, C.~Keeler, V.~Lysov and A.~Strominger,
  %``Wilsonian Approach to Fluid/Gravity Duality,''
\href{http://dx.doi.org/10.1007/JHEP03(2011)141}  {JHEP {\bf 1103}, 141 (2011)}
[\href{http://arxiv.org/abs/arXiv:1006.1902}{arXiv:1006.1902 [hep-th]}].
  %%CITATION = ARXIV:1006.1902;%%
  %83 citations counted in INSPIRE as of 02 Sep 2013

%\cite{Bredberg:2011jq}
\bibitem{Bredberg:2011jq}
  I.~Bredberg, C.~Keeler, V.~Lysov and A.~Strominger,
  %``From Navier-Stokes To Einstein,''
  \href{http://dx.doi.org/10.1007/JHEP07(2012)146}{JHEP {\bf 1207}, 146 (2012)}
  [\href{http://arxiv.org/abs/arXiv:1101.2451}{arXiv:1101.2451 [hep-th]}].
  %%CITATION = ARXIV:1101.2451;%%
  %53 citations counted in INSPIRE as of 02 Sep 2013

%\cite{Bredberg:2011xw}
\bibitem{Bredberg:2011xw}
  I.~Bredberg and A.~Strominger,
  %``Black Holes as Incompressible Fluids on the Sphere,''
  \href{http://dx.doi.org/10.1007/JHEP05(2012)043}{JHEP {\bf 1205}, 043 (2012)}
  [\href{http://arxiv.org/abs/arXiv:1106.3084}{arXiv:1106.3084 [hep-th]}].
  %%CITATION = ARXIV:1106.3084;%%
  %15 citations counted in INSPIRE as of 03 Sep 2013

%\cite{Anninos:2011zn}
\bibitem{Anninos:2011zn}
  D.~Anninos, T.~Anous, I.~Bredberg and G.~S.~Ng,
  %``Incompressible Fluids of the de Sitter Horizon and Beyond,''
  \href{http://dx.doi.org/10.1007/JHEP05(2012)107}{JHEP {\bf 1205}, 107 (2012)}
  [\href{http://arxiv.org/abs/arXiv:1110.3792}{arXiv:1110.3792 [hep-th]}].
  %%CITATION = ARXIV:1110.3792;%%
  %7 citations counted in INSPIRE as of 03 Sep 2013


%\cite{Compere:2011dx}
\bibitem{Compere:2011dx}
  G.~Compere, P.~McFadden, K.~Skenderis and M.~Taylor,
  %``The Holographic fluid dual to vacuum Einstein gravity,''
  \href{http://dx.doi.org/10.1007/JHEP07(2012)146}{JHEP {\bf 1107}, 050 (2011)}
  [\href{http://arxiv.org/abs/arXiv:1103.3022}{arXiv:1103.3022 [hep-th]}].
  %%CITATION = ARXIV:1103.3022;%%
  %45 citations counted in INSPIRE as of 02 Sep 2013

%\cite{Compere:2012mt}
\bibitem{Compere:2012mt}
  G.~Compere, P.~McFadden, K.~Skenderis and M.~Taylor,
  %``The relativistic fluid dual to vacuum Einstein gravity,''
  \href{http://dx.doi.org/10.1007/JHEP03(2012)076}{JHEP {\bf 1203}, 076 (2012)}
  [\href{http://arxiv.org/abs/arXiv:1201.2678}{arXiv:1201.2678 [hep-th]}].
  %%CITATION = ARXIV:1201.2678;%%
  %18 citations counted in INSPIRE as of 03 Sep 2013

%\cite{Eling:2012ni}
\bibitem{Eling:2012ni}
  C.~Eling, A.~Meyer and Y.~Oz,
  %``The Relativistic Rindler Hydrodynamics,''
  \href{http://dx.doi.org/10.1007/JHEP05(2012)116}{JHEP {\bf 1205}, 116 (2012)}
  [\href{http://arxiv.org/abs/arXiv:1201.2705}{arXiv:1201.2705 [hep-th]}].
  %%CITATION = ARXIV:1201.2705;%%
  %14 citations counted in INSPIRE as of 07 Sep 2013

%\cite{Meyer:2013sva}
\bibitem{Meyer:2013sva}
  A.~Meyer and Y.~Oz,
  %``Constraints on Rindler Hydrodynamics,''
  \href{http://dx.doi.org/10.1007/JHEP07(2013)090}{JHEP {\bf 1307}, 090 (2013)}
  [\href{http://arxiv.org/abs/arXiv:1304.6305}{arXiv:1304.6305}].
  %%CITATION = ARXIV:1304.6305;%%


%AdS solution
%

%\cite{Lysov:2011xx}
\bibitem{Lysov:2011xx}
  V.~Lysov and A.~Strominger,
  %``From Petrov-Einstein to Navier-Stokes,''
  \href{http://arxiv.org/abs/arXiv:1104.5502}{arXiv:1104.5502 [hep-th]}.
  %%CITATION = ARXIV:1104.5502;%%
  %21 citations counted in INSPIRE as of 03 Sep 2013

%\cite{Huang:2011he}
\bibitem{Huang:2011he}
  T.~-Z.~Huang, Y.~Ling, W.~-J.~Pan, Y.~Tian and X.~-N.~Wu,
  %``From Petrov-Einstein to Navier-Stokes in Spatially Curved Spacetime,''
  \href{http://dx.doi.org/10.1007/JHEP10(2011)079}{JHEP {\bf 1110}, 079 (2011)}
  [\href{http://arxiv.org/abs/arXiv:1107.1464}{arXiv:1107.1464 [gr-qc]}].
  %%CITATION = ARXIV:1107.1464;%%
  %13 citations counted in INSPIRE as of 03 Sep 2013

%\cite{Huang:2011kj}
\bibitem{Huang:2011kj}
  T.~-Z.~Huang, Y.~Ling, W.~-J.~Pan, Y.~Tian and X.~-N.~Wu,
  %``Fluid/gravity duality with Petrov-like boundary condition in a spacetime with a cosmological constant,''
  \href{http://dx.doi.org/10.1103/PhysRevD.85.123531}{Phys.\ Rev.\ D {\bf 85}, 123531 (2012)}
  [\href{http://arxiv.org/abs/arXiv:1111.1576}{arXiv:1111.1576 [hep-th]}].
  %%CITATION = ARXIV:1111.1576;%%
  %8 citations counted in INSPIRE as of 03 Sep 2013

%\cite{Zhang:2012uy}
\bibitem{Zhang:2012uy}
  C.~-Y.~Zhang, Y.~Ling, C.~Niu, Y.~Tian and X.~-N.~Wu,
  %``Magnetohydrodynamics from gravity,''
  \href{http://dx.doi.org/10.1103/PhysRevD.86.084043}{Phys.\ Rev.\ D {\bf 86}, 084043 (2012)}
  [\href{http://arxiv.org/abs/arXiv:1204.0959}{arXiv:1204.0959 [hep-th]}].
  %%CITATION = ARXIV:1204.0959;%%
  %7 citations counted in INSPIRE as of 03 Sep 2013


%\cite{Wu:2013kqa}
\bibitem{Wu:2013kqa}
  X.~Wu, Y.~Ling, Y.~Tian and C.~Zhang,
  %``Fluid/Gravity Correspondence for General Non-rotating Black Holes,''
  \href{http://dx.doi.org/10.1088/0264-9381/30/14/145012}{Class.\ Quant.\ Grav.\  {\bf 30}, 145012 (2013)}
  [\href{http://arxiv.org/abs/arXiv:1303.3736}{arXiv:1303.3736 [hep-th]}].
  %%CITATION = ARXIV:1303.3736;%%
  %2 citations counted in INSPIRE as of 03 Sep 2013

%\cite{Wu:2013mda}
\bibitem{Wu:2013mda}
  B.~Wu and L.~Zhao,
  %``Gravity-mediated holography in fluid dynamics,''
  \href{http://dx.doi.org/10.1016/j.nuclphysb.2013.05.017}{Nucl.\ Phys.\ B {\bf 874}, 177 (2013)}
  [\href{http://arxiv.org/abs/arXiv:1303.4475}{arXiv:1303.4475 [hep-th]}].
  %%CITATION = ARXIV:1303.4475;%%
  %1 citations counted in INSPIRE as of 03 Sep 2013


%\cite{Ling:2013kua}
\bibitem{Ling:2013kua}
  Y.~Ling, C.~Niu, Y.~Tian, X.~N.~Wu and W.~Zhang,
  %``The Petrov-like boundary condition at finite cutoff surface in Gravity/Fluid duality,''
  \href{http://dx.doi.org/10.1103/PhysRevD.90.043525}{Phys.\ Rev.\ D {\bf 90}, 043525 (2014)}
  [\href{http://arxiv.org/abs/arXiv:1306.5633}{arXiv:1306.5633 [gr-qc]}].
  %%CITATION = ARXIV:1306.5633;%%
  %6 citations counted in INSPIRE as of 28 Aug 2014


%\cite{Lysov:2013jsa}
\bibitem{Lysov:2013jsa}
  V.~Lysov,
  %``On the Magnetohydrodynamics/Gravity Correspondence,''
  \href{http://arxiv.org/abs/arXiv:1310.4181}{arXiv:1310.4181 [hep-th]}.
  %%CITATION = ARXIV:1310.4181;%%

%\cite{Wu:2014fwa}
\bibitem{Wu:2014fwa}
  B.~Wu and L.~Zhao,
  %``Holographic fluid from the nonminimally coupled scalar¨Ctensor theory of gravity,''
  \href{http://dx.doi.org/10.1088/0264-9381/31/10/105018}{Class.\ Quant.\ Grav.\  {\bf 31}, 105018 (2014)}
  [\href{http://arxiv.org/abs/arXiv:1401.6487}{arXiv:1401.6487 [hep-th]}].
  %%CITATION = ARXIV:1401.6487;%%

%\cite{Cai:2013uye}
\bibitem{Cai:2013uye}
  R.~-G.~Cai, L.~Li, Q.~Yang and Y.~-L.~Zhang,
  %``Petrov type $I$ Condition and Dual Fluid Dynamics,''
  \href{http://dx.doi.org/10.1007/JHEP04(2013)118}{JHEP {\bf 1304}, 118 (2013)}
  [\href{http://arxiv.org/abs/arXiv:1302.2016}{arXiv:1302.2016 [hep-th]}].
  %%CITATION = ARXIV:1302.2016;%%
  %2 citations counted in INSPIRE as of 03 Sep 2013



%\bibitem{exac}
%E.~Hertl, C.~Hoenselaers, D.~Kramer, M. Maccallum, and H. Stephani,
%%``Exact solutions of Einstein's field equations''
%[\href{http://dx.doi.org/10.1017/CBO9780511535185}
%{Cambridge University Press, 2nd, 2003}]
  %\cite{Coley:2004jv}

%\cite{Coley:2004jv}
\bibitem{Coley:2004jv}
  A.~Coley, R.~Milson, V.~Pravda and A.~Pravdova,
  %``Classification of the Weyl tensor in higher dimensions,''
  \href{http://dx.doi.org/10.1088/0264-9381/21/7/L01}{Class.\ Quant.\ Grav.\  {\bf 21}, L35 (2004)}
  [\href{http://arxiv.org/abs/gr-qc/0401008}{gr-qc/0401008}].
  %%CITATION = GR-QC/0401008;%%
  %120 citations counted in INSPIRE as of 04 Sep 2013


%\cite{Coley:2007tp}
\bibitem{Coley:2007tp}
  A.~Coley,
  %``Classification of the Weyl Tensor in Higher Dimensions and Applications,''
  \href{http://dx.doi.org/10.1088/0264-9381/25/3/033001}{Class.\ Quant.\ Grav.\  {\bf 25}, 033001 (2008)}
  [\href{http://arxiv.org/abs/arXiv:0710.1598}{arXiv:0710.1598 [gr-qc]}].
  %%CITATION = ARXIV:0710.1598;%%
  %59 citations counted in INSPIRE as of 04 Sep 2013


%Petrov¡ª¡ª¡ª¡ª¡ª¡ª¡ª¡ª¡ª¡ª¡ª¡ª¡ª¡ª¡ª¡ª¡ª¡ª¡ª¡ª¡ª¡ª¡ª¡ª¡ª¡ª¡ª¡ª¡ª¡ª¡ª¡ª¡ª¡ª¡ª¡ª¡ª¡ª¡ª¡ª¡ª¡ª¡ª¡ª¡ª¡ª¡ª¡ª¡ª¡ª¡ª¡ª¡ª¡ª¡ª¡ª¡ª¡ª¡ª¡ª¡ª¡ª¡ª¡ª¡ª¡ª¡ª¡ª¡ª¡ª¡ª¡ª¡ª¡ª¡ª¡ª¡ª¡ª¡ª¡ª¡ª¡ª¡ª¡ª¡ª¡ª¡ª¡ª¡ª¡ª¡ª¡ª¡ª¡ª¡ª¡ª¡ª¡ª¡ª¡ª
%%Petrov¡ª¡ª¡ª¡ª¡ª¡ª¡ª¡ª¡ª¡ª¡ª¡ª¡ª¡ª¡ª¡ª¡ª¡ª¡ª¡ª¡ª¡ª¡ª¡ª¡ª¡ª¡ª¡ª¡ª¡ª¡ª¡ª¡ª¡ª¡ª¡ª¡ª¡ª¡ª¡ª¡ª¡ª¡ª¡ª¡ª¡ª¡ª¡ª¡ª¡ª¡ª¡ª¡ª¡ª¡ª¡ª¡ª¡ª¡ª¡ª¡ª¡ª¡ª¡ª¡ª¡ª¡ª¡ª¡ª¡ª¡ª¡ª¡ª¡ª¡ª¡ª¡ª¡ª¡ª¡ª¡ª¡ª¡ª¡ª¡ª¡ª¡ª¡ª¡ª¡ª¡ª¡ª¡ª¡ª¡ª¡ª¡ª¡ª¡ª¡ª
%%Petrov¡ª¡ª¡ª¡ª¡ª¡ª¡ª¡ª¡ª¡ª¡ª¡ª¡ª¡ª¡ª¡ª¡ª¡ª¡ª¡ª¡ª¡ª¡ª¡ª¡ª¡ª¡ª¡ª¡ª¡ª¡ª¡ª¡ª¡ª¡ª¡ª¡ª¡ª¡ª¡ª¡ª¡ª¡ª¡ª¡ª¡ª¡ª¡ª¡ª¡ª¡ª¡ª¡ª¡ª¡ª¡ª¡ª¡ª¡ª¡ª¡ª¡ª¡ª¡ª¡ª¡ª¡ª¡ª¡ª¡ª¡ª¡ª¡ª¡ª¡ª¡ª¡ª¡ª¡ª¡ª¡ª¡ª¡ª¡ª¡ª¡ª¡ª¡ª¡ª¡ª¡ª¡ª¡ª¡ª¡ª¡ª¡ª¡ª¡ª¡ª
%
%\bibitem{petr}
%A.~Petrov, %``Einstein Spaces'', Pergamon Press, 1969.

%%%
%Gauss-Nonnet
%%
%\cite{Chirco:2011ex}
\bibitem{Chirco:2011ex}
  G.~Chirco, C.~Eling and S.~Liberati,
  %``Higher Curvature Gravity and the Holographic fluid dual to flat spacetime,''
\href{http://dx.doi.org/10.1007/JHEP08(2011)009}{JHEP {\bf 1108}, 009 (2011)}
[\href{http://arxiv.org/abs/arXiv:1105.4482}{arXiv:1105.4482 [hep-th]}].
  %%CITATION = ARXIV:1105.4482;%%
  %8 citations counted in INSPIRE as of 02 Sep 2013

%\cite{Eling:2012xa}
\bibitem{Eling:2012xa}
  C.~Eling, A.~Meyer and Y.~Oz,
  %``Local Entropy Current in Higher Curvature Gravity and Rindler Hydrodynamics,''
  \href{http://dx.doi.org/10.1007/JHEP08(2012)088}{JHEP {\bf 1208}, 088 (2012)}
  [\href{http://arxiv.org/abs/arXiv:1205.4249}{arXiv:1205.4249 [hep-th]}].
  %%CITATION = ARXIV:1205.4249;%%
  %3 citations counted in INSPIRE as of 22 Jul 2013

%\cite{Cai:2014sua}
\bibitem{Cai:2014sua}
  R.~G.~Cai, Q.~Yang and Y.~L.~Zhang,
  %``Petrov type I Condition and Rindler Fluid in Vacuum Einstein-Gauss-Bonnet Gravity,''
  \href{http://arxiv.org/abs/arXiv:1408.6488}{arXiv:1408.6488 [hep-th]}.
  %%CITATION = ARXIV:1408.6488;%%




%Fluid/gravity¡ª¡ª¡ª¡ª¡ª¡ª¡ª¡ª¡ª¡ª¡ª¡ª¡ª¡ª¡ª¡ª¡ª¡ª¡ª¡ª¡ª¡ª¡ª¡ª¡ª¡ª¡ª¡ª¡ª¡ª¡ª¡ª¡ª¡ª¡ª¡ª¡ª¡ª¡ª¡ª¡ª¡ª¡ª¡ª¡ª¡ª¡ª¡ª¡ª¡ª¡ª¡ª¡ª¡ª¡ª¡ª¡ª¡ª¡ª¡ª¡ª¡ª¡ª¡ª¡ª¡ª¡ª¡ª¡ª¡ª¡ª¡ª¡ª¡ª¡ª¡ª¡ª¡ª¡ª¡ª¡ª¡ª¡ª¡ª¡ª¡ª¡ª¡ª¡ª¡ª¡ª¡ª¡ª¡ª¡ª¡ª
%%Fluid/gravity¡ª¡ª¡ª¡ª¡ª¡ª¡ª¡ª¡ª¡ª¡ª¡ª¡ª¡ª¡ª¡ª¡ª¡ª¡ª¡ª¡ª¡ª¡ª¡ª¡ª¡ª¡ª¡ª¡ª¡ª¡ª¡ª¡ª¡ª¡ª¡ª¡ª¡ª¡ª¡ª¡ª¡ª¡ª¡ª¡ª¡ª¡ª¡ª¡ª¡ª¡ª¡ª¡ª¡ª¡ª¡ª¡ª¡ª¡ª¡ª¡ª¡ª¡ª¡ª¡ª¡ª¡ª¡ª¡ª¡ª¡ª¡ª¡ª¡ª¡ª¡ª¡ª¡ª¡ª¡ª¡ª¡ª¡ª¡ª¡ª¡ª¡ª¡ª¡ª¡ª¡ª¡ª¡ª¡ª¡ª¡ª
%%%Fluid/gravity¡ª¡ª¡ª¡ª¡ª¡ª¡ª¡ª¡ª¡ª¡ª¡ª¡ª¡ª¡ª¡ª¡ª¡ª¡ª¡ª¡ª¡ª¡ª¡ª¡ª¡ª¡ª¡ª¡ª¡ª¡ª¡ª¡ª¡ª¡ª¡ª¡ª¡ª¡ª¡ª¡ª¡ª¡ª¡ª¡ª¡ª¡ª¡ª¡ª¡ª¡ª¡ª¡ª¡ª¡ª¡ª¡ª¡ª¡ª¡ª¡ª¡ª¡ª¡ª¡ª¡ª¡ª¡ª¡ª¡ª¡ª¡ª¡ª¡ª¡ª¡ª¡ª¡ª¡ª¡ª¡ª¡ª¡ª¡ª¡ª¡ª¡ª¡ª¡ª¡ª¡ª¡ª¡ª¡ª¡ª¡ª
%


%\cite{Cai:2011xv}
\bibitem{Cai:2011xv}
  R.~-G.~Cai, L.~Li and Y.~-L.~Zhang,
  %``Non-Relativistic Fluid Dual to Asymptotically AdS Gravity at Finite Cutoff Surface,''
  \href{http://dx.doi.org/10.1007/JHEP07(2011)027}{JHEP {\bf 1107}, 027 (2011)}
  [\href{http://arxiv.org/abs/arXiv:1104.3281}{arXiv:1104.3281 [hep-th]}].
  %%CITATION = ARXIV:1104.3281;%%
  %23 citations counted in INSPIRE as of 03 Sep 2013

%\cite{Kuperstein:2011fn}
\bibitem{Kuperstein:2011fn}
  S.~Kuperstein and A.~Mukhopadhyay,
  %``The unconditional RG flow of the relativistic holographic fluid,''
  \href{http://dx.doi.org/10.1007/JHEP11(2011)130}{JHEP {\bf 1111}, 130 (2011)}
  [\href{http://arxiv.org/abs/arXiv:1105.4530}{arXiv:1105.4530 [hep-th]}].
  %%CITATION = ARXIV:1105.4530;%%
  %12 citations counted in INSPIRE as of 07 Sep 2013

%\cite{Brattan:2011my}
\bibitem{Brattan:2011my}
  D.~Brattan, J.~Camps, R.~Loganayagam and M.~Rangamani,
  %``CFT dual of the AdS Dirichlet problem : Fluid/Gravity on cut-off surfaces,''
  \href{http://dx.doi.org/10.1007/JHEP12(2011)090}{JHEP {\bf 1112}, 090 (2011)}
  [\href{http://arxiv.org/abs/arXiv:1106.2577}{arXiv:1106.2577 [hep-th]}].
  %%CITATION = ARXIV:1106.2577;%%
  %24 citations counted in INSPIRE as of 05 Sep 2013

%\cite{Matsuo:2012pi}
\bibitem{Matsuo:2012pi}
  Y.~Matsuo, M.~Natsuume, M.~Ohta and T.~Okamura,
  %``The Incompressible Rindler fluid versus the Schwarzschild-AdS fluid,''
  \href{http://dx.doi.org/10.1093/ptep/pts069}{PTEP {\bf 2013}, 023B01 (2013)}
  [\href{http://arxiv.org/abs/arXiv:1206.6924}{arXiv:1206.6924 [hep-th]}].
  %%CITATION = ARXIV:1206.6924;%%
  %5 citations counted in INSPIRE as of 05 Sep 2013

%\cite{Kuperstein:2013hqa}
\bibitem{Kuperstein:2013hqa}
  S.~Kuperstein and A.~Mukhopadhyay,
  %``Spacetime emergence via holographic RG flow from incompressible Navier-Stokes at the horizon,''
  \href{http://dx.doi.org/10.1007/JHEP11(2013)086}{JHEP {\bf 1311}, 086 (2013)}
  [\href{http://arxiv.org/abs/arXiv:1307.1367}{arXiv:1307.1367 [hep-th]}].
  %%CITATION = ARXIV:1307.1367;%%
  %5 citations counted in INSPIRE as of 14 Feb 2014


%\cite{CYZ}
\bibitem{CYZ}
  R.~-G.~Cai, Q.~Yang and Y.~-L.~Zhang,
  %``Petrov type \I Condition and Rindler Hydrodynamics in Higher Curvature Gravity,''
 to appear.



%\cite{Caldarelli:2012hy}
\bibitem{Caldarelli:2012hy}
  M.~M.~Caldarelli, J.~Camps, B.~Gouteraux and K.~Skenderis,
  %``AdS/Ricci-flat correspondence and the Gregory-Laflamme instability,''
  \href{http://dx.doi.org/10.1103/PhysRevD.87.061502}{Phys.\ Rev.\ D {\bf 87}, 061502 (2013)}
  [\href{http://arxiv.org/abs/arXiv:1211.2815}{arXiv:1211.2815 [hep-th]}].
  %%CITATION = ARXIV:1211.2815;%%
  %11 citations counted in INSPIRE as of 05 Sep 2013

%\cite{Caldarelli:2013aaa}
\bibitem{Caldarelli:2013aaa}
  M.~M.~Caldarelli, J.~Camps, B.~Gout¨¦raux and K.~Skenderis,
  %``AdS/Ricci-flat correspondence,''
  \href{http://dx.doi.org/10.1007/JHEP04(2014)071}{JHEP {\bf 1404}, 071 (2014)}
  [\href{http://arxiv.org/abs/arXiv:1312.7874}{arXiv:1312.7874 [hep-th]}].
  %%CITATION = ARXIV:1312.7874;%%
  %1 citations counted in INSPIRE as of 28 Jan 2014

%\cite{Cai:2012vr}
\bibitem{Cai:2012vr}
  R.~-G.~Cai, L.~Li, Z.~-Y.~Nie and Y.~-L.~Zhang,
  %``Holographic Forced Fluid Dynamics in Non-relativistic Limit,''
  \href{http://dx.doi.org/10.1016/j.nuclphysb.2012.06.014}{Nucl.\ Phys.\ B {\bf 864}, 260 (2012)}
  [\href{http://arxiv.org/abs/arXiv:1202.4091}{arXiv:arXiv:1202.4091 [hep-th]}].
  %%CITATION = ARXIV:1202.4091;%%
  %14 citations counted in INSPIRE as of 28 Jan 2014

%\cite{Pinzani-Fokeeva:2014cka}
\bibitem{Pinzani-Fokeeva:2014cka}
  N.~Pinzani-Fokeeva and M.~Taylor,
  %``Towards a general fluid/gravity correspondence,''
  \href{http://arxiv.org/abs/arXiv:1401.5975}{arXiv:1401.5975 [hep-th]}.
  %%CITATION = ARXIV:1401.5975;%%


\end{thebibliography}
\end{document}